# Efficient Cybersecurity Assessment Using SVM and Fuzzy Evidential Reasoning for Resilient Infrastructure

Zaydon L. Ali[1], Wassan Saad Abduljabbar Hayale[2], Israa Ibraheem Al_Barazanchi[3], Ravi Sekhar[4*], Pritesh Shah[4], Sushma Parihar[4]

[1] College of Political Science, Mustansiriyah University, Baghdad 10001, Iraq
[2] Electrical Engineering Department, Engineering College, Aliraqia University, Baghdad 10001, Iraq
[3] Computer Technology Engineering Department, College of Information Technology, Imam Ja'afar Al-Sadiq University, Baghdad 10001, Iraq
[4] Symbiosis Institute of Technology (SIT) Pune Campus, Symbiosis International (Deemed University) (SIU), Pune 412115, Maharashtra, India

Corresponding Author Email: ravi.sekhar@sitpune.edu.in

**ABSTRACT**

With current advancement in hybermedia knowledges, the privacy of digital information has developed a critical problem. To overawed the susceptibilities of present security protocols, scholars tend to focus mainly on efforts on alternation of current protocols. Over past decade, various proposed encoding models have been shown insecurity, leading to main threats against significant data. Utilizing the suitable encryption model is very vital means of guard against various such, but algorithm is selected based on the dependency of data which need to be secured. Moreover, testing potentiality of the security assessment one by one to identify the best choice can take a vital time for processing. For faster and precise identification of assessment algorithm, we suggest a security phase exposure model for cipher encryption technique by invoking Support Vector Machine (SVM). In this work, we form a dataset using usual security components like contrast, homogeneity. To overcome the uncertainty in analysing the security and lack of ability of processing data to a risk assessment mechanism. To overcome with such complications, this paper proposes an assessment model for security issues using fuzzy evidential reasoning (ER) approaches. Significantly, the model can be utilised to process and assemble risk assessment data on various aspects in systematic ways. To estimate the performance of our framework, we have various analyses like, recall, F1 score and accuracy.

## 1. INTRODUCTION

The mix of machine learning and big data is becoming increasingly common as artificial intelligence advances. It has several uses in e-commerce, banking, shipping, medical and health care, and other fields. Machine learning will be used everywhere as a result of the widespread promotion of big data technologies and its integration with the Internet of Things. However, enormous amounts of data will undoubtedly generate privacy issues during the storage, conversation, and application processes. Machine learning facility providers, for example, have gained access to data provided by users throughout the education and prediction phases and may readily get private information, leading in privacy breaches. As a classic machine learning approach, SVM is used to tackle numerous categorization issues and was initially utilised by Bell Laboratories to identify handwritten digital libraries [1]. SVM has drawn the attention of researchers in different domains and has been extensively advocated as a result of its amazing effect. SVM now has various uses in computer vision [2, 3], medical diagnostics [4, 5], and other fields. SVM plays a vital part in the progress of artificial intelligence because of its capacity to address multidimensional information and complex feature issues and integrate categorization of internal maximization with a kernel approach created on theory of statistical learning. In comparison to other predictive procedures, SVM overcomes the "overlearning" issues in a narrow sample size and corrects the local peak fault. Now, secure multiparty addition (MPC) and homomorphic encryption (HE) are the key research skills for privacy-preserving SVM employing encryption techniques. MPC contains a considerable quantity of data conversation, which cannot match the real efficiency criteria in SVM. At HE schemes, completely HE systems have poor effectiveness and remain at the theoretical experiment stage in SVM. Schemes that utilise incomplete HE is the mainstays for meeting practical needs. However, these partly homomorphic methods all have three flaws. The first drawback is that ciphertext computation is inefficient. Some HE systems must alter the required SVM formulae in order to realize sophisticated ciphertext computation, which not only enhances the quantity of ciphertext analysis but also enhances the quantity of data communication among operators and servers. Another drawback is the inability to scale. SVM is classified into two types: linear SVM and quadratic SVM. Furthermore, various kernel functions are accessible for nonlinear SVM. The majority of extant partial HE techniques are tailored to a particular kind of SVM. The schemes must be modified when the kind of SVM changes. As a result, the capacity of these solutions is weak. The third drawback is that users spend a lot of time online. Because certain methods need cooperative computing between users and servers, users must remain connected. Our technique enables efficient HE calculations and equivalent prediction accuracy with an unencrypted scenario while imposing a low computational cost on a regulator.

The training phase is the process of building supervised ML classifiers (e.g., SVM), which trains a given classifier from a collection of considered data. It has been demonstrated that the accuracy of ML classifiers improves with increasing direction of the quantity of training information. Because a single entity's training dataset (e.g., a hospital or a group of providers) is typically constrained in relationships of data quantity and diversity, there has long been a demand for an effective technique to train ML classifiers utilising a mix of datasets acquired from many entities.

## 2. LITERATURE SURVEY

Because the creation of ML models typically involves many parties, privacy aim is to train the perfect with data from everyone involved while preventing the information of a particular gathering from being learnt by other revelries. This study fits under this group, which has had a plethora of solutions presented over the previous decade [6].

### 2.1 Privacy

Differential privacy is a popular strategy for protecting data privacy during the publishing process. DP maintains security of released data by introducing precisely designed perturbations into the original information. The study in literature [7] introduced a deep learning strategy based on DP that allows several participants to cooperatively construct a neural network while safeguarding confidential data in their datasets. Because all designs are conducted over plaintext statistics, DP-based systems can attain significant computational competence. However, because the perturbations always diminish the quality of the input data, the resultant ML models may be erroneous. Furthermore, agitations may not completely guarantee data confidentiality because only a small quantity of private knowledge about each unique exercise data is disclosed. In DP, for example, a privacy budget parameter is used to balance data privacy with efficiency of the model: A greater budget protects privacy of the data while decreasing accuracy of the methodology. In directive to improve security of information in ML exercise, HC is used to train machine learning representations on converted data. The fundamental properties of HC enable calculations on ciphertext while retaining its accuracy. Several secure HC-based algorithms, including as SVM [8], logistic analysis [9], decision trees [10], and Naive Bayes [11], have been developed for training various ML models. González-Serrano et al. [12] built a private trained SVM algorithm and procedures for safe multiplication and division based on Paillier. Because Paillier does not support some computations, they utilize the server authorization, an important third group, for compute subcontracting. In comparison to DP-based explanations, HC-based solutions provide improved privacy of data at the expense of low competence. There are two explanations for this.

1. Though fully homomorphic encryption (FHE) allows for complex calculations (e.g., with arbitrary improvements) on ciphertexts, present FHE applications are prohibitively expensive in view of encryption and processing, providing them with impractical in real world scenario.
2. Although partly homomorphic encryption (PHE) is more applied than FHE, it only enables one kind of operation (addition or multiplication). As a result, present solutions often rely on an established external (e.g., the authorization site) to enable complicated computations or lead to erroneous models by approximating complicated formulas with a single kind of computation.

### 2.2 ML classification for privacy preservation

A sample information for categorization and the ML perfect are often owned by two distinct gatherings in a classification-as-a-service situation. The data owner is eager to learn the classification outcome but is hesitant to provide a sample with sample information to an untrustworthy perfect owner in order to acquire classification service. Meanwhile, the model operator may be hesitant to divulge the categorization model's information because it is an extremely important resource to the service provider. Efforts have been made to build effective ways to preserve both parties' privacy. The study in literature [13] suggested a multilayer learning-based technique for categorising encrypted pictures. They rely on the notion that the copy information should be kept private. The study in literature [14] suggested an online medical pre diagnosis confidentiality nonlinear SVM classification technique. Both the confidential data in everyone's medical history and the SVM model are safeguarded by their enterprise.

Yu et al. [15] proposed a robust privacy SVM data classification model applicable to multiclass evaluations. This model ensures that the server does not gain any knowledge about the input information from the clients, and it also conceals the server-side classifier from the users throughout the classification process. There are many studies that have utilized HC methodologies to establish a range of classification processes. These processes make use of typical ML classifiers such as decision trees, selection of hyperplanes, and Naive Bayes, all operating on encrypted data.

Xie et al. [16] introduced the inaugural method of classifying data while preserving privacy, utilizing SVM on horizontally split data. This involved breaking down the core function issue of horizontally divided data into individual function issues for each data section. The advantage of this is that it can safeguard local data and hide the classification model by generating the local Gramme medium separately for each party and then compiling the global prototype through MPC. However, a significant drawback is that it requires at least three participants to conduct a closed-loop operational sequence, which is both complex and inefficient. In 2008, Vaidya and colleagues put forward a method for preserving privacy in SVMs that divide data. They managed to secure the global kernel vector using HE without revealing the original data. While this method has improved security, it has unfortunately led to a decrease in efficiency.

Khalaf et al. [13] suggested a novel approach that enhances on Vaidya's technique. The arbitrary matrix is introduced as the disturbance amount in the scheme, and the local information is provided to a genuine third party to compute the worldwide SVM method, which prevents the possibility of data leaking produced by the revelation of the local Gramme matrix and has superior security. Though, when there are many players, the computational complexity increases dramatically, as does the communication cost. Shibghatullah et al. [14] introduced a privacy-conserving SVM built around the function of the Gaussian kernel in 2014, which was the first time the Paillier encryption technique was used to SVM. However, Rahulamathavan's system necessitates several data linkages and is inefficient. Then, several individuals enhanced on Rahulamathavan's plan.

Yu et al. [15] introduced a nonlinear SVM-based privacy-securing model for diagnostic system. Although Zhu's system has been improved in terms of efficiency, it is not mountable enough. Xie et al. [16] enhanced the safety and other features of Rahulamathavan's plan, but the core structure of the organization was not modified, and its competence was not increased. Al-Barazanchi et al. [17] suggested a privacy online diagnostic approach leveraging subcontracted SVM, as well as safe procedures and associated data technique based on the Paillier cryptosystem. Their approach not only secures user-supplied data and diagnosis findings, but it also prevents the evaluation model from being accessible by cloud servers and users [18-25].

## 3. SUPPORT VECTOR MACHINE

SVMs are a form of supervised learning algorithms employed for both regression and classification tasks [26-28]. They are counted among the broad family of generalized linear classification. The unique attributes of SVMs are their capacity to reduce empirical error in classification and the geometric boundary [29-33]. Due to these features, SVMs are often referred to as Maximum margin classifiers.

The foundation of SVMs lies in the principle of Structural Risk Minimization (SRM). This principle enables SVMs to transform the original vector into a multidimensional space where a large separating hyperplane gets constructed. On either side of this separating hyperplane that partitions the data, two parallel hyperplanes are established. The separation hyperplane is characterized by its superior distance from the two parallel areas [34].

The methodology which employs the maximum distance or margin between various hyperplanes results in a stronger classifier's generalization error. This means that the greater the distance between these hyperplanes, the lower the likelihood of making errors during the generalization process. Thus, SVMs provide a robust and efficient tool for machine learning tasks.

We include data points of the for

$$\{(a_1, b_1), (a_2, b_2), (a_3, b_3) \ldots (a_m, b_m)\} \quad (1)$$

where, $b_m$=1/-1, a continuous denoting parameter to which the information belongs. M depicts the total samples. Every $a_m$ is the real parameter [35]. The scaling is significant to protect against larger variables. We could view this data in the training data, by providing the hyperplane which takes

$$v.a + x = 0 \quad (2)$$

where, $x$ represents the scalar and $v$ is the vector which is in various dimension. The vector $v$ is perpendicular to the dividing hyperplane. Additionally, the parameter offset $x$ provides to enhance the margin. In the absence of $x$, the hyperplane is enhanced to pass over the origin avoiding the solution to the problem, we are illustrating the utilization of SVM parallel hyperplane [36].

$$v.a + x = 1 \quad (3)$$

$$v.a + x = -1 \quad (4)$$

If the simulated data is linearly distinct, we can choose these hyperplanes with no points in between and then strive to maximise their distance.

A splitting hyperplane with the greatest specified margin by N=2/|V| that is significantly support the vectors data points to it.

$$b_i[V^T.a_i + y] = 1 \quad (5)$$

The Optimal Canonical Hyperplane (OCH) is a canonical Hyperplane with the greatest margin. OCH should meet the following conditions for all data.

$$b_i[V^T.a_i + y] \geq 1 \quad (6)$$

where, $l$ is the total amount of training data points. To select the best separating hyperplane with the greatest margin, a learning model should lower the constraint of inequality.

### 3.1 SVM kernel selection

Training parameter $a_i$ are plotted into maximum dimensional space by the utilization of component α. Then SVM identifies a linear dividing hyperplane with higher margin in this maximum space of dimensionality. B>0 is the penalty component of the error function.

Moreover, $F(a_i, a_j) = \alpha(a_i)^T \alpha(a_i)$ is the known function. Because SVM has numerous kernel functions, how to choose a decent kernel feature is also a study topic. However, there are certain common kernel functions for general use:

- Kernel function: $F(a_i, a_j) = \alpha(a_i)^T$
- Kernel polynomial function: $F(a_i, a_j) = (\pi \alpha(a_i)^T + r)^d, \pi>0$
- RBF function: $F(a_i, a_j) = exp(\pi \alpha(a_i)^T + r)^d, \pi>0$
- Sigmoid function: $F(a_i, a_j) = \tanh(\pi \alpha(a_i)^T + r)^d$

where, α, π and d are the kernel components. In these functions used in the SVM, RBF is the significant function used due to following reasons:

1. Samples are mapped using the RBF kernel functions.
2. The RBF function has lower hyperparameters than the kernel function.
3. Numerical difficulties are lower in the kernel functions used.

| Algorithm 1 |
|---|
| **Input:** Kernel matrix, class parameter $b \in \{-1,1\}^m$ and the margin component D |
| **Output:** weighted vector $\pi \in X_+^m$. |
| Function of kernel (G, b, D) |
| Step 1: $\pi \leftarrow 0$ |
| Step 2: repeat |
| Step 3:     for j=1,……, m do |
| Step 4:         $\pi_j \leftarrow \pi_j + b_j \cdot (1 - b_j \cdot \sum_{i=1}^m g, \pi, b_j)$ |
| Step 5:     if $\pi_j > D$ then $\pi_j \leftarrow D$ |
| Step 6:         if $\pi_j < 0$ then $\pi_j \leftarrow 0$ |
| Step 7:     end for |
| Step 8: until conjunction |
| End of function |

## 4. FUZZY REASONING METHODOLOGY

The normal credibility mass score of the estimation index $f_i$ is

$$n_i\{P_m\} = \alpha \beta_m \tag{7}$$

$$n_i\{H\} = 1 - \sum_{m=1}^{M} n_i\{P_m\} \tag{8}$$

Then $n_i\{P_m\}$ is the normal reliability function of the estimation index $f_i$ at the estimation level of $P_m$; $n_i\{P_m\}$ is the reliability which is not assigned, classifying the degree to which the proof that has not been allocated being created. Let

$$\bar{n}_i\{P\} = 1 - \alpha_i \tag{9}$$

The N estimation indexes consists in the estimation object are associated with proof. The certain methodology is as follows

$$n_i - N\{P_m\} = j\{\prod_{i=1}^{N}[n_i\{P_m\} + n_i\{P_m\}] \\ - \prod_{i=1}^{N} n_i\{H\}\}, n = 1,2, \dots, M \tag{10}$$

$$n_i - N\{P_m\} = j\{\prod_{i=1}^{N}\{n_i\{P\}\} \tag{11}$$

$$\bar{n}_{1-N}\{P\} = j\{\prod_{i=1}^{N}\{n_i\{P\}\} \tag{12}$$

The overall risk assessment mass score $n_i - N\{P_m\}$ of the object evaluated which is obtained giving to fusion operations, $n_i - N\{P_{m+1}\}$ is the mass score at the association $P_{m,m+1}$ of the fuzzy estimation phase.

## 5. RESULT AND DISCUSSION

By examining the graph, you can easily see how the framework performs at each epoch. The blue and red lines depict training and validation accuracy, respectively. The model's accuracy and loss are then shown. Typically, validation and training efficiency improves over time. The accuracy of the model after fitting it is 0.6954, and it has not yet been exposed to training. Essentially, when testing passes validation, you should stop training since the model is starting to recall specific items rather than learning patterns. In this example, validation accuracy continues to improve even after the sixth epoch. The loss is minimised using the RNN method via back propagation with auto differentiation. The model's performance demonstrates that the loss decreases with each epoch, from 0.22 to 0.0314 for the first and sixth epochs, respectively. Figure 1 illustrates the training and validation loss and Figure 2 depicts the accuracy obtained during training and validation.

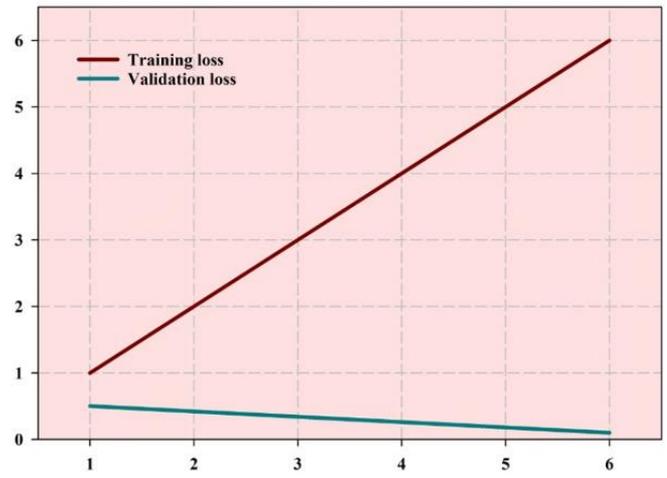

**Figure 1.** Loss during training and validation

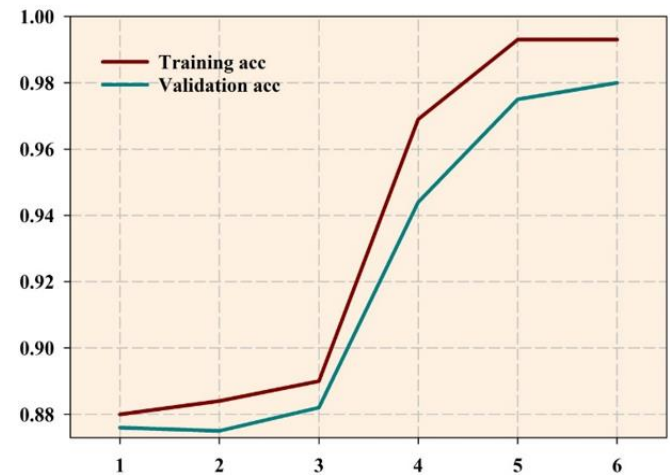

**Figure 2.** Accuracy during training and validation

The model created using deep learning was then contrasted with the SVM model to determine which model was better suited for the classification of binary data. Other researchers' previous studies have validated SVM as the best match for SMS categorization. We trained, tested, and evaluated the SVM model in order to compare them. We utilised a same dataset, sample amount, random state, and all other parameters remained unchanged. The validation precision and loss of SVM model classifier utilising the same dataset parameters. According to the figure, SVM has a slightly better accuracy in validation and training of 0.96 compared with navies 0.92, while SVM has a significantly lower false positive rate.

Figure 3 illustrates the computational overhead of fuzzy evidential reasoning approaches where the number of iterations for estimating the computational overhead is shown for Heart Disease Data Set (HDDS). Proposed ER model outperformed the existing naïve model as the number of iterations keeps on increasing. During the 100th iteration the overhead is 890 when compared to the proposed model which is 650. Figure 4 illustrates the computation overhead for Breast Cancer Data Set (BCDS), where the computational overhead of the proposed model is 78% of computational overhead whereas the existing method overhead is about 89%.

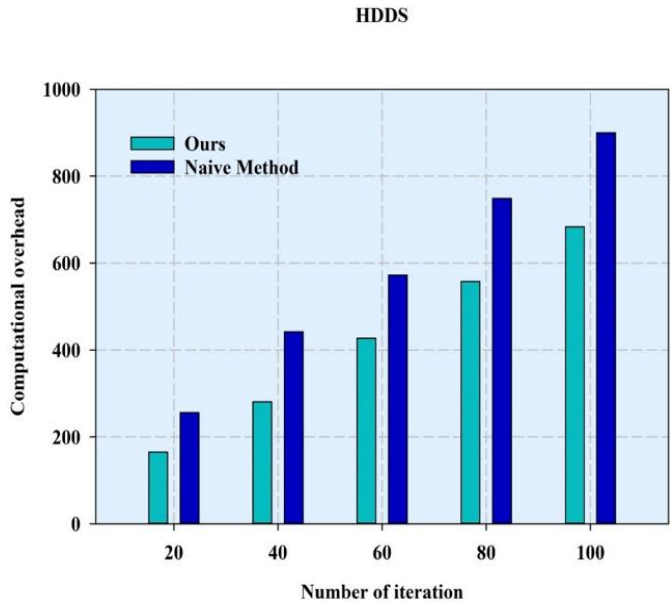

**Figure 3.** Computational overhead of ER and Naïve method for HDDS

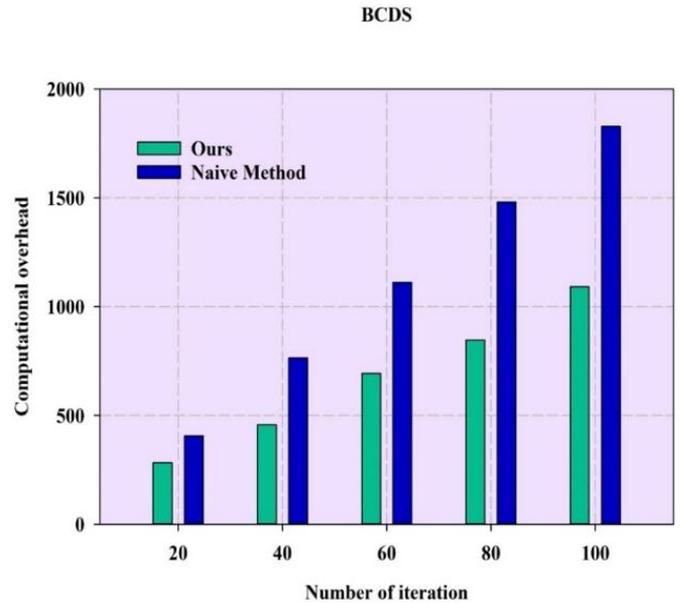

**Figure 4.** Computational overhead of ER and Naïve method for BCDS

**Table 1.** Classification performance

|      | Model | Training Set | Testing Set | Recall | Precision |
|------|-------|--------------|-------------|--------|-----------|
| HDDS | Proposed (SVM-ER) | 230 | 62 | 83.70% | 85.45% |
|      | Navie method | 230 | 62 | 81.80% | 82.30% |
| BCDS | Proposed (SVM-ER) | 455 | 114 | 94.20% | 92.15% |
|      | Navie method | 455 | 114 | 91.10% | 89.20% |
| IDS  | Proposed (SVM-ER) | 263 | 80 | 76.40% | 91.40% |
|      | Navie method | 263 | 80 | 74.76% | 89.20% |

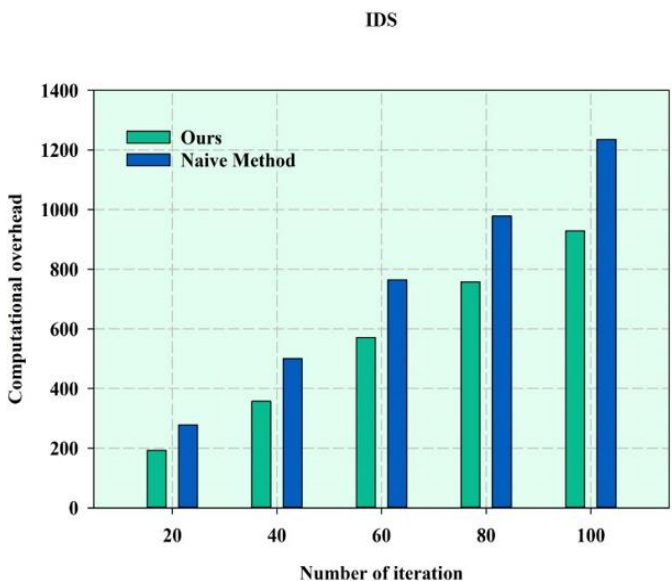

**Figure 5.** Computational overhead of ER and naïve method for IDS

Table 1 illustrates the overall computational performance on various dataset which is utilised. Aside from privacy protection, some prior work has explored model prediction verification. That is, after getting the forecasts from the model forecasting service provider, the user confirms their accuracy. The verification technique must verify that the user cannot get the model's key parameters stored by the service provider. Figure 5 illustrates the computational overhead of the Ionosphere Data Set (IDS) obtained from Proposed ER and the existing Naïve method. To use these strategies in our system, we must convert a significant number of variables in the model used for SVM development to integers, which may result in truncation errors. To assess the influence of these mistakes on our scheme's classification performance, an appropriate comparison is the classic SVM technique without regard for privacy protection.

## 6. CONCLUSION

In this paper, the overcome the susceptibilities of the existing security protocols, for faster and precisive identification of assessment algorithm, we suggest a security phase detection model for cipher encryption technique by invoking Support Vector Machine. this paper proposes an assessment model for security issues using fuzzy evidential reasoning (ER) approach. We utilised a same dataset, sample amount, random state, and all other parameters remained unchanged. SVM has a slightly better accuracy in validation and training of 0.96 compared with navies 0.92, while SVM has a significantly lower false positive rate. The validation precision and loss of SVM model classifier utilising the same dataset parameters. The model's performance demonstrates that the loss decreases with each epoch, from 0.22 to 0.0314 for the first and sixth epochs. The computational overhead of the proposed model is 78% of computational overhead whereas the existing method overhead is about 89%.